\documentclass[structabstract]{aa}  
\usepackage{graphicx}
\usepackage{txfonts}
\usepackage{natbib}
\bibpunct{(}{)}{;}{a}{}{,}
\usepackage{amssymb}
\usepackage{xcolor}
\usepackage{lipsum}
\usepackage{textcomp}
\usepackage[colorlinks=true,linkcolor=blue,citecolor=blue,
bookmarks=true,bookmarksopen=true,bookmarksnumbered=true,draft=false]{hyperref}
\usepackage{enumerate}

\newcommand{\rsun}{$R_{\sun}$}

\newcommand{\LXP}{LXP 169}

\newcommand{\xmm}{{\it XMM-Newton}}
\newcommand{\swift}{{\it Swift}}

\newcommand{\hour}{$^{\mathrm{h}}$}
\newcommand{\minute}{$^{\mathrm{m}}$}
\newcommand{\second}{$^{\mathrm{s}}$}

\begin{document}

   \title{Discovery of a 168.8 s X-ray pulsar transiting in front of its\\ Be
companion star in the Large Magellanic Cloud\,\thanks{Based on observations
obtained with \xmm, an ESA science mission with instruments and contributions
directly funded by ESA Member States and NASA.}}

   \subtitle{}

	\titlerunning{Discovery of a 168.8 s X-ray pulsar transiting in front of
its Be companion star in the LMC}

   \author{P. Maggi    \inst{1}
	\and F.   Haberl   \inst{1}
	\and R.   Sturm    \inst{1}
	\and W.   Pietsch  \inst{1}
	\and A.   Rau      \inst{1}
	\and J.   Greiner  \inst{1}
	\and A.   Udalski  \inst{2}
	\and M.    Sasaki \inst{3}
	}

   \institute{Max-Planck-Institut f\"ur extraterrestrische Physik, Postfach
	1312, Giessenbachstr., 85741 Garching, Germany\\ \email{pmaggi@mpe.mpg.de}
	\and
	Warsaw University Observatory, Al. Ujazdowskie 4, 00-478 Warzawa, Poland
	\and
	Institut f\"ur Astronomie und Astrophysik T\"ubingen, Universit\"at
	T\"ubingen, Sand 1, 72076 T\"ubingen, Germany
    }

   \date{Received 5 February 2013\,/\,Accepted 5 April 2013}

  \abstract
  {}
	{
We report the discovery of LXP 169, a new high-mass X-ray binary in the Large
Magellanic Cloud. The optical counterpart has been identified and appears to
exhibit an eclipsing light curve. We performed follow-up observations to clarify
the eclipsing nature of the system.}
	{
Energy spectra and time series were extracted from two \xmm\ observations to
search for pulsations, characterise the spectrum, and measure spectral and
timing changes. Long-term X-ray variability was studied using archival
\textit{ROSAT} data. The \xmm\ positions were used to identify the optical
counterpart. We
obtained ultraviolet to near-infrared photometry to characterise the companion,
along with its 4000\,d $I$-band light curve and colour-magnitude variability. We
observed LXP 169 with \swift\ at two predicted eclipse times.}
	{
We found a spin period of 168.8 s that did not change between two \xmm\
observations. The X-ray spectrum, well characterised by a power law, was harder
when the source was brighter. The X-ray flux of LXP 169 is found to be variable
by a factor of at least 10. The counterpart is highly variable on short and long
timescales, and its photometry is that of an early-type star with an ouflowing
circumstellar disc producing a near-infrared excess. This classifies the source
as a Be/X-ray binary pulsar. We observe a transit in the ultraviolet, thereby
confirming that the companion star itself is eclipsed. We give an ephemeris for
the transit of MJD $56203.877_{-0.197}^{+0.934} + N \times (24.329\pm0.008)$. We
propose and discuss the scenario where the matter captured from the companion's
equatorial disc creates an extended region of high density around the neutron
star, which partially eclipses the companion as the neutron star transits in
front of it.}
	{This is most likely the first time the compact object in an X-ray binary
is observed to eclipse its companion star. \LXP\ would be the first eclipsing
Be/X-ray binary, and a wealth of important information might be gained from
additional observations, such as a measure of the possible Be
disc\,/\,orbital plane misalignment, or the mass of the neutron star.
}

   \keywords{X-rays: binaries -- Stars: neutron -- Stars: emission-line, Be --
binaries: eclipsing -- X-rays: individuals: LXP 169 -- Magellanic Clouds} 

   \maketitle

\begin{table*}[t]
\caption{Details of the X-ray observations}
\label{table_info}
\centering
\begin{tabular}{@{}c@{\hspace{0.13cm}} @{\hspace{0.13cm}}c@{\hspace{0.13cm}}
@{\hspace{0.13cm}}c@{\hspace{0.13cm}} @{\hspace{0.13cm}}c@{\hspace{0.13cm}}
@{\hspace{0.13cm}}c@{\hspace{0.13cm}} @{\hspace{0.13cm}}c@{\hspace{0.13cm}}
@{\hspace{0.13cm}}c@{\hspace{0.13cm}} @{\hspace{0.13cm}}c@{\hspace{0.13cm}}
@{\hspace{0.13cm}}c@{\hspace{0.13cm}} @{\hspace{0.13cm}}c@{}}
\hline\hline
\noalign{\smallskip}
Identifier & Date of observation start & \multicolumn{2}{c}{Central coordinates
(J2000)}
& \multicolumn{3}{c}{Total\,/\,filtered exposure time (ks) \tablefootmark{a}} &
Off-axis & Orbital \\
 & (date\,/\,MJD)  & RA & DEC &  pn & MOS1 & MOS2 & angle \tablefootmark{b} &
phase \tablefootmark{d} \\
\noalign{\smallskip}
\hline
\noalign{\smallskip}
 & & \multicolumn{2}{c}{\xmm} & & & & & \\
\noalign{\smallskip}
\hline
\noalign{\smallskip}
0690742501 (XMM1) & 2012 Jul 13\,/\,56121.23 & 05\hour\,06\minute\,40.9\second
& $-$68\degr\,18\minute\,30\second &  24.2/19.4 & 26.7/20.1 &
26.7/20.2 & 9.5 & 0.603-0.612\\
0690742401 (XMM2) & 2012 Sep 09\,/\,56179.06 & 05\hour\,09\minute\,44.3\second &
$-$68\degr\,21\minute\,13\second & 29.4/26.0 & 30.5/27.2 & 30.5/27.2
& 10.8 & 0.980--0.994 \\
\noalign{\smallskip}
\hline
\noalign{\smallskip}
 & & \multicolumn{2}{c}{\textit{ROSAT}} & \multicolumn{2}{c}{Exposure time}  &
Count rate & Off-axis & Orbital \\
& & & & \multicolumn{2}{c}{(ks)} & ($10^{-3}$ cts\,s$^{-1}$) &
angle\tablefootmark{b} & phase \\
\noalign{\smallskip}
\hline
\noalign{\smallskip}
rp300129n00 & 1992 Apr 09\,/\,48721.07 & 05\hour\,08\minute\,00.0\second
& $-$68\degr\,37\minute\,48\second & \multicolumn{2}{c}{3.7} & 
3.2$\pm$1.2 & 12.8  & 0.433--0.447 \\
rp500037n00 & 1992 Apr 09\,/\,48721.54 & 04\hour\,49\minute\,00.0\second
& $-$68\degr\,43\minute\,48\second & \multicolumn{2}{c}{4.8} & 
8.2$\pm$1.6 & 19.7 & 0.452--0.618 \\
rp500060n00 & 1992 Apr 11\,/\,48723.20 & 05\hour\,05\minute\,43.2\second
& $-$67\degr\,52\minute\,47\second & \multicolumn{2}{c}{3.9} &
$< 3.1$ \tablefootmark{c} & 34.8  & 0.520--0.662 \\
\noalign{\smallskip}
\hline
\noalign{\smallskip}
 & & \multicolumn{2}{c}{\swift/XRT} & \multicolumn{2}{c}{}  & & & \\
\noalign{\smallskip}
\hline
\noalign{\smallskip}
00032578001 & 2012 Oct 02\,/\,56202.00 & 05\hour\,07\minute\,51.3\second
& $-$68\degr\,24\minute\,25\second & \multicolumn{2}{c}{1.1} & $<$7.1 
\tablefootmark{c} & ---  & 0.923--0.942\\
00032578002 & 2012 Oct 03\,/\,56203.61 & 05\hour\,07\minute\,49.2\second
& $-$68\degr\,24\minute\,20\second & \multicolumn{2}{c}{3.5} & $<$4.0
\tablefootmark{c} & ---  & 0.989--1.001 \\
00032578003 & 2012 Oct 04\,/\,56204.81 & 05\hour\,07\minute\,55.4\second
& $-$68\degr\,24\minute\,45\second & \multicolumn{2}{c}{1.0} & $<$8.8
\tablefootmark{c} & ---  & 0.038--0.039 \\
00032578004 & 2012 Oct 10\,/\,56210.41 & 05\hour\,07\minute\,37.5\second
& $-$68\degr\,21\minute\,33\second & \multicolumn{2}{c}{1.3} & $<$11.4
\tablefootmark{c} & ---  & 0.269--0.275 \\
00032578005 & 2012 Oct 27\,/\,56227.71 & 05\hour\,07\minute\,50.6\second
& $-$68\degr\,23\minute\,59\second & \multicolumn{2}{c}{1.0} & 12.5$\pm$4.0 &
---  & 0.979--0.980 \\
00032578006 & 2012 Oct 28\,/\,56228.25 & 05\hour\,07\minute\,56.4\second
& $-$68\degr\,24\minute\,18\second & \multicolumn{2}{c}{5.2} & 9.6$\pm$1.6 &
---  & 0.002--0.013 \\
00032578007 & 2012 Oct 29\,/\,56229.51 & 05\hour\,07\minute\,52.6\second
& $-$68\degr\,23\minute\,22\second & \multicolumn{2}{c}{0.3} & $<$50.4
\tablefootmark{c} & ---  & 0.054 \\
\hline
\end{tabular}
\tablefoot{
\tablefoottext{a}{Performed duration (total) and useful (filtered) exposure
times, after removing high background intervals.}
\tablefoottext{b}{Angle in arcmin between the central coordinates and the X-ray
source. \swift\ observations were on-axis.}
\tablefoottext{c}{$3\sigma$ upper limit.}
\tablefoottext{a}{Using ephemeris given in Eq.\,\ref{eq_ephemeris}.}
}
\end{table*}

\section{Introduction}
\label{introduction}

Be/X-ray binaries (hereafter BeXRBs) are a major subclass of X-ray binaries. In
these systems, a compact object accretes material from a normal companion star.
The optical counterparts are non-supergiant, emission-line stars, which have
spectral classes later than O5 and earlier than B9, with the bulk of the
population concentrated around B0--B1 \citep{2005MNRAS.356..502C}.
Classical OBe stars are rapid rotators surrounded by an equatorial disc
of circumstellar material. The disc emits lines, chiefly the Balmer and Paschen
series of hydrogen, but also a few He and Fe lines. An infrared excess is also
produced by the equatorial disc \citep[for a recent review, see
e.g.][]{2011Ap&SS.332....1R}.

In BeXRBs the neutron star (there is no confirmed black hole/Be X-ray binary
yet) is usually in a wide orbit with a significant eccentricity
\citep[orbital periods of tens to a few hundred of days, and $0.3 \lesssim e
\lesssim 0.9$,][]{2011MNRAS.416.1556T} around its companion, leading to a
transient nature of the system. Copious amounts of X-rays can be produced when
the neutron star captures material from the equatorial disc of the Be star. This
occurs when the separation between the two components is the smallest, i.e. at
or near periastron passage, and leads to the so-called Type I X-ray outbursts,
which last for a small fraction of the orbital period and have X-ray
luminosities $L_X \sim 10^{36-37}$ erg\,s$^{-1}$. Less frequently, giant (Type
II) outbursts can occur, reaching luminosities in excess of $10^{37}$
erg\,s$^{-1}$ and lasting for several orbital periods. Although many questions
remain open, it has been suggested that giant outbursts are associated to
warping episodes of a Be disc misaligned with respect to the orbital plane
\citep[see][and references therein]{2012arXiv1211.5225O}.

The study of BeXRBs in our nearest neighbour galaxies, the Large and Small
Magellanic Clouds (LMC and SMC, respectively), is rewarding. It presents several
advantages over a similar study in our own Galaxy (in particular, low
foreground absorption and known distances). In recent years a large population
of BeXRBs has been identified in the SMC, with a total of $\sim 60$ confirmed
systems \citep{2004A&A...414..667H,2005MNRAS.356..502C,2010ASPC..422..224C}.
About 45 candidates have also been identified during the \xmm\ survey of the SMC
\citep{2013A&A...000A..00S}. While the LMC is about ten times as massive as the
SMC, it contains only 14 confirmed BeXRBs so far
\citep{2005A&A...442.1135L,2006ATel..783....1M,2012A&A...542A.109S,
2013MNRAS.428.3607K}. This discrepancy is possibly explained by different star
formation histories (SFHs). \citet{2010ApJ...716L.140A} find that the locations
of SMC BeXRBs correlate with stellar populations of ages $\sim$25--60 Myr.
Despite large spatial variations, the most recent episodes of enhanced star
formation activity in the LMC occurred 12 Myr and 100 Myr ago
\citep{2009AJ....138.1243H}. This is different from the time at which
most Be stars develop their equatorial discs, which was found to peak at $\sim
40$ Myr \citep{2005ApJS..161..118M}. However, the X-ray coverage of the LMC is
still not as complete as for the SMC, precluding early interpretations on the
role of  different SFHs. The situation will improve in upcoming years, with the
completion of an \xmm\ survey, homogeneously covering the $\sim$ 4.5\degr
$\times$ 4.5\degr\ central region of the LMC (Haberl et al., in prep.). Even
though no single-epoch survey can reveal the complete population of transient
BeXRBs, we are likely to detect many new systems.

In this work we present the discovery of a new member of the BeXRB population
of the LMC, recently identified in our survey with \xmm. The X-ray observations
and results are described in Sect.\,\ref{xray}. The analysis of the optical
companion is presented in Sect.\,\ref{optical_counterpart}. The system presents
unusual and very interesting properties, which we analyse and discuss in
Sects.\,\ref{eclipses} \& \ref{results_discussion}.
Our findings are then summarised in Sect.\,\ref{summary}.

\section{X-ray analysis}
\label{xray}

	\subsection{\xmm\ observations and data reduction}
	\label{xray_xmm_observations}
In the course of the ongoing \xmm\ survey of the LMC, we discovered a bright
source in an observation (ObsId 0690742501, hereafter XMM1) carried out on 2012
July 13. In a later observation (ObsId 0690742401, hereafter XMM2) on 2012
September 9, the source was again detected. The European Photon Imaging Camera
(EPIC), comprising a pn CCD imaging camera \citep{2001A&A...365L..18S} and two
MOS CCD imaging cameras \citep{2001A&A...365L..27T}, was used as the primary
instrument. Both observations were performed using the same instrumental
setting: the cameras were operated in full-frame mode, with the thin and medium
optical filter for pn and MOS cameras, respectively. We used the XMM
SAS\,\footnote{Science Analysis Software, \url{http://xmm.esac.esa.int/sas/}}
version 11.0.1 for the data reduction. In observation XMM1, the source had a
0.2--12 keV EPIC-pn count rate of $0.610\pm0.017$ cts\,s$^{-1}$
(background-subtracted and corrected for vignetting and the finite point spread
function), while it was fainter in the second observation ($0.105\pm0.004$
cts\,s$^{-1}$).

The first observation was affected by a high background. The useful exposure
time for the pn detector was only $\sim$ 5.8~ks. Because the source is bright,
the effect of the background is less important, and we increased the pn rate
threshold for background filtering to improve the photon statistics of the
spectra and time series. Instead of the standard thresholds of 8
cts\,ks$^{-1}$\,arcmin$^{-2}$ (in the 7--15 keV band), which we apply to image
analyses, we used a higher value of 50 cts\,ks$^{-1}$\,arcmin$^{-2}$. Doing so,
we reached an useful exposure time of $\sim$ 19~ks. The MOS cameras were less
affected and we used a 7--15 keV rate threshold for background filtering of 2.5
cts\,ks$^{-1}$ arcmin$^{-2}$, which yielded 20~ks of useful exposure times.
Observation XMM2 was less contaminated by solar flares. Using rate thresholds of
8 and 2.5 cts\,ks$^{-1}$ arcmin$^{-2}$ for pn and MOS yielded exposure times in
excess of 26~ks. In Table \ref{table_info} we list the details of all the X-ray
observations.

\subsection{\xmm\ timing analysis}
	\label{xray_results_timing}
\begin{figure}[t]
	\centering
	\includegraphics[bb= 66 46 556 700, clip,angle=-90,width=0.999\hsize]
{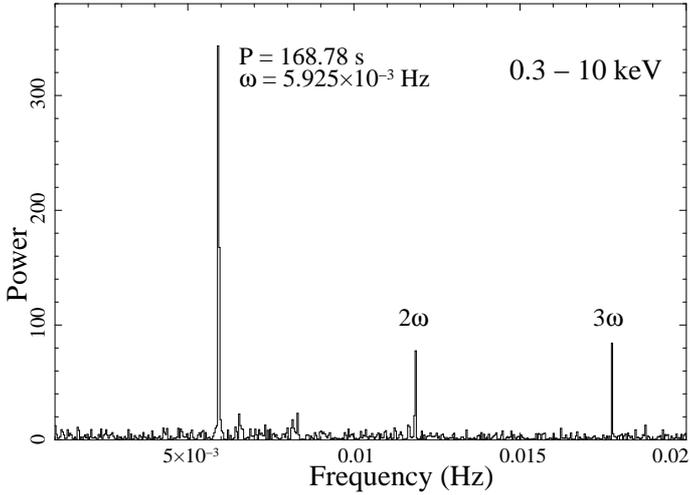}
	\caption{Power spectrum density of \LXP\ for observation XMM1, combining all
EPIC cameras in the 0.3 -- 10 keV energy range. The prominent power peak at
$\omega=5.925$ mHz and its second and third harmonics are labelled.}
\label{fig_power_spectrum}
\end{figure}

We produced time series (barycentre corrected) for observations XMM1 and XMM2.
For the timing analysis we used the merged time series, combining pn and MOS
data across the 0.3--10 keV range in common Good Times Intervals. The power
spectrum density of observation XMM1 (shown in Fig.\,\ref{fig_power_spectrum})
prominently peaks at $\omega_{\mathrm{peak}} = 5.925 \times 10^{-3}$ Hz. The
second and third harmonics are detected as well.

We used the Bayesian periodic signal detection method described in
\citet{1996ApJ...473.1059G} to determine the pulse period $P_s$ with a $1\sigma$
uncertainty. For observation XMM1 we found $P_s = 168.777 \pm 0.006$ s. During
the XMM2 observation we measured $P_s = 168.788 \pm 0.042$ s. The time series
from observation XMM2 contained less events (1918) than the time series of
observation XMM1 (13329 events), when the source was brighter, and this explains
the (purely statistical) larger uncertainty measured in the second observation.
Within the uncertainties, no variation of $P_s$ is detected.

\begin{figure}[t]
	\centering
	\includegraphics[width=1.0\hsize]
{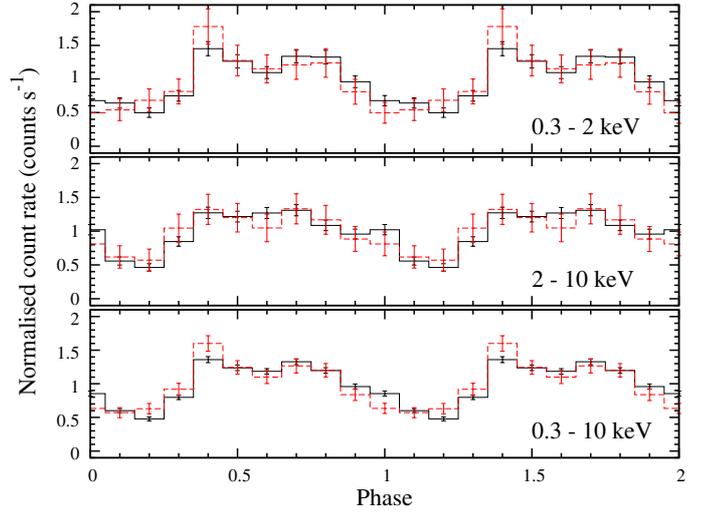}
	\caption{X-ray pulse profile of \LXP\ as observed in observation XMM1 (solid
black line) and XMM2 (red dashed line), combining all EPIC cameras. The profiles
are shown in different energy bands (labelled), and have been
background-subtracted. Each profile has been normalised to its respective
average net count rate.}
\label{fig_pulse_profile}
\end{figure}

In Fig.\,\ref{fig_pulse_profile}, we show the background-corrected X-ray pulse
profiles folded to the period measured from observation XMM1 (having lower
statistical uncertainty). Within the error bars, no significant profile
differences are seen between observation XMM1 and XMM2. In the broad 0.3--10 keV
band we estimate pulse fractions of 39.7$\pm$2.2 per cent and
38.9$_{-5.1}^{+4.4}$ per cent in the XMM1 and XMM2 datasets, respectively,
assuming a sinusoidal pulse profile. Thus, the pulse fraction shows no
variability between the two epochs.

In line with the detection of pulsations and due to the confusion in the
\textit{ROSAT} nomenclature (see Sect.\,\ref{xray_rosat_observations}), we
propose \object{LXP 169} as the identifier for this source. The acronym ``LXP''
stands for ``Large Magellanic Cloud X-ray Pulsar'' and the following number
indicates the spin period of the system to three significant figures. We thereby
follow the notation of \citet{2005MNRAS.356..502C} who used ``SXP'' for X-ray
pulsars in the SMC. In the remainder of the paper we refer to the source as
\LXP.

\subsection{X-ray position}
	\label{xray_position}
The statistical $1\sigma$ uncertainty in the position of \LXP\ for observation
XMM1 and XMM2 is $r_{stat}=0.17$\arcsec\ and 0.24\arcsec, respectively, to
which a systematic uncertainty $r_{syst}$, accounting for the uncertainty in the
spacecraft pointing, should be added. $r_{syst}$ is typically in the range
1\arcsec--2\arcsec\ \citep{2009A&A...493..339W} but can be improved to $\sim
0.5$\arcsec\ by correlating the X-ray sources detected in an observation with
optical counterparts having a better astrometry. As finding the true optical
counterpart is crucial, it is definitely worthwhile to improve the precision of
the X-ray position.

In the field of view of our two observations there were no bright, previously
known sources to use. Instead, we correlated the catalogue of
mid-infrared-selected active galactic nucleii (AGN) candidates from
\citet{2009ApJ...701..508K} with our X-ray detection lists. While these sources
have not yet been spectroscopically confirmed \citep[only a fraction of them
have been already observed, see][]{2012ApJ...746...27K} and risks of
contamination exist, we checked that the hardness ratios of the X-ray sources
with a mid-IR counterpart were consistent with an AGN nature. Note that
\citet{2012ApJ...746...27K} showed that the false positive rate is significantly
reduced in the presence of an X-ray source. The positions in the catalogue of
\citet{2009ApJ...701..508K} are from the Spitzer survey of the LMC
\citep{2006AJ....132.2268M}, matched to the 2MASS catalogue, which have a
typical uncertainty of 0.3\arcsec\ (1$\sigma$). We used 0.8\arcsec\ as a
systematic uncertainty in the corrected positions to take this into account.

We identified three and seven X-ray source\,/\.mid-IR AGN candidate associations
in observations XMM1 and XMM2, respectively, which we used to compute the
average right ascension and declination offsets. The total linear shifts are
2.4\arcsec\ and 2.3\arcsec. The corrected X-ray positions of \LXP\ are listed
in Table~\ref{table_spectral}.

\begin{figure*}[t]
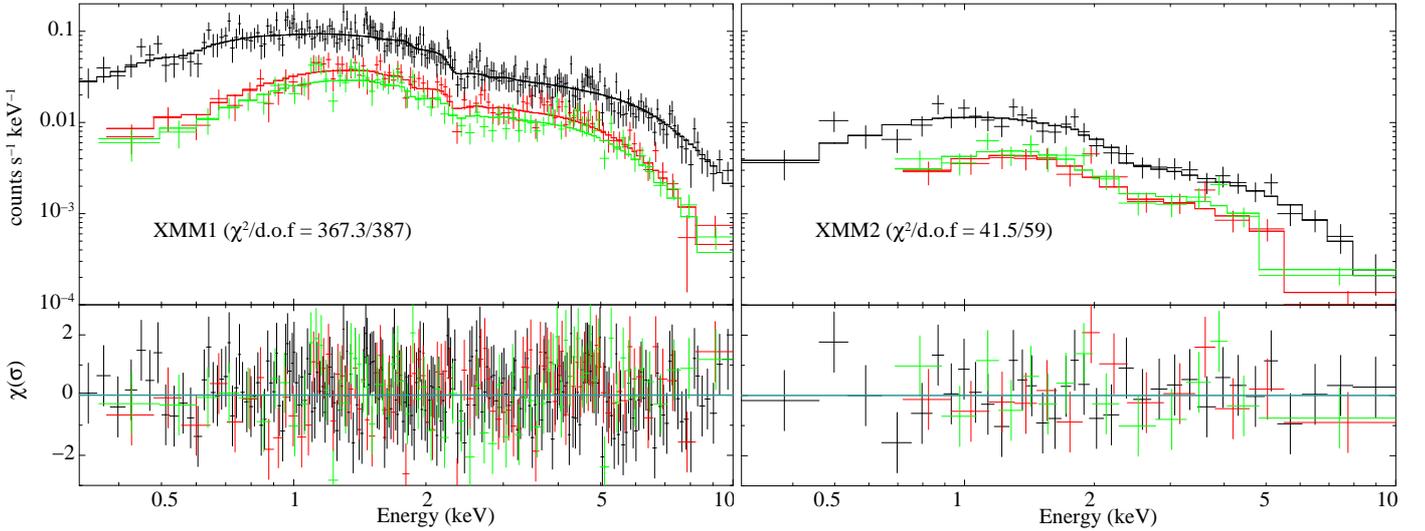

\centering
\includegraphics[bb= 80 26 580 699 ,clip,angle=-90,width=0.52\hsize]
{spectrum_XMM1.ps}
\includegraphics[bb= 80 92 580 701 ,clip,angle=-90,width=0.471\hsize]
{spectrum_XMM2.ps}
\caption{EPIC spectrum of \LXP\ for observation XMM1 (left) and XMM2 (right).
The black, red, and green points are data from pn, MOS1, and MOS2 cameras,
respectively. The histograms show the folded best-fit model (absorbed
power law). Residuals are shown in the lower panels in terms of $\sigma$.
}
\label{fig_xray_spectra}
\end{figure*}

\subsection{Spectral analysis}
\label{xray_results_spectral}
We extracted energy spectra from observations XMM1 and XMM2 for all three
cameras. Source and background spectra were taken from circles with radii of
22.5\arcsec\ and 42.5\arcsec, respectively. Single and double-pixel events
(\texttt{PATTERN} = 0 to 4) from the pn detector were used and all single to
quadruple-pixel (\texttt{PATTERN} = 0 to 12) events from the MOS detectors were
used. We rebinned the spectra with a minimum of 20 counts per bin in order to
allow the use of the $\chi ^2$-statistic. The spectral analysis was performed
with XSPEC \citep{1996ASPC..101...17A} version 12.7.0u.

We fitted the EPIC spectra in the 0.3--10 keV range with an absorbed power law.
Two photoelectric absorption components were used, one with solar abundances
\citep[according to the table of][]{2000ApJ...542..914W} and a fixed column
density N$_{H\mathrm{\ Gal}} = 5.93 \times 10^{20}\ \mathrm{cm}^{-2}$ for the
Galactic foreground absorption \citep{1990ARA&A..28..215D}, and another one with
a free N$_{H\mathrm{\ LMC}}$ for the LMC. Metal abundances for the latter were
fixed to the average metallicity in the LMC \citep[\emph{i.e.} half their solar
values,][]{1992ApJ...384..508R}.

\begin{figure}[t]
	\centering
	\includegraphics[bb= 72 46 575 700, clip,angle=-90,width=0.89\hsize]
{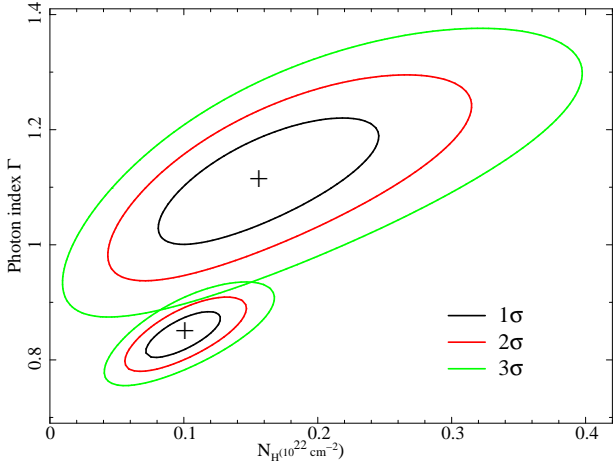}
	\caption{$\Gamma$\,--\,N$_H$ confidence plane for observation XMM1 (smaller
$\Gamma$) and XMM2 (larger $\Gamma$ and errors). 1, 2, and 3$\sigma$ contours
are shown. The best-fit values are marked by the black crosses.}
\label{fig_confidence}
\end{figure}

Spectra from the three EPIC cameras were fitted simultaneously and we included
cross-calibration factors for the MOS data ($C_{MOS1}$ and $C_{MOS2}$). The
absorbed power law provided satisfactory fits, as seen from the $\chi^2$ and
residuals (spectra are shown in Fig.\,\ref{fig_xray_spectra}). No additional
component, such as a black-body one, was needed. The best-fit parameter values
are given in Table\,\ref{table_spectral}, along with the derived fluxes and
luminosities. Bad pixels and bad columns in the extraction regions explain the
behaviour of $C_{MOS1}$ and $C_{MOS2}$. In XMM2, the source was partially
located on a pn CCD gap, resulting in $\sim 20$\% higher $C_{MOS1}$ and
$C_{MOS2}$. Consequently, the flux quoted for the second observation comes from
MOS data.

We show the confidence regions in the $\Gamma$\,--\,N$_H$ plane for observations
XMM1 and XMM2 in Fig.\,\ref{fig_confidence}. The spectrum is harder in the first
observation, when the source was about six times brighter. The column density
is poorly constrained in the (softer and fainter) second
observation, precluding a comparison of N$_H$ between the two epochs. We note
that the anticorrelation of the spectral index and the X-ray flux (that is, the
brighter the source, the harder the spectrum) has been observed in other
high-mass (likely Be/) X-ray binary pulsars in outburst
\citep[e.g.][]{2002ApJ...569..903B,2009MNRAS.395.1662I}. This behaviour is
consistent with the source being in the so-called ``horizontal branch''
\citep[as defined by][]{2008A&A...489..725R,2012arXiv1212.5944R}, traced by
BeXRBs in the hardness-intensity diagram during giant outbursts, as long as they
are below a critical luminosity.

\begin{table}[t]
\caption{\xmm\ X-ray results.}
\label{table_spectral}
\centering
\begin{tabular}{l c c }
\hline
\hline
\noalign{\smallskip}
 Parameter & XMM1 & XMM2 \\
\noalign{\smallskip}
\hline
\noalign{\smallskip}
RA (J2000) & 05\hour\,07\minute\,55.38\second & 05\hour\,07\minute\,55.25\second
\\
DEC (J2000) & $-$68\degr\,25\minute\,04.6\second &
$-$68\degr\,25\minute\,06.0\second \\
\noalign{\smallskip}
$P_{s}$ (s) & $ 168.777 \pm 0.006$ & $168.788 \pm 0.042$\\
\noalign{\smallskip}
N$_{H\mathrm{\ LMC}}$ ($10^{20}\ \mathrm{cm}^{-2}$) & 9.84$_{-2.9}^{+3.2}$ &
15.4$_{-7.8}^{+10.1}$ \\
Photon index $\Gamma$ & 0.84$\pm0.04$ & 1.11$\pm0.12$ \\
$C_{MOS1}$ & 1.06$\pm0.05$ &  1.17$_{-0.15}^{+0.16}$ \\
$C_{MOS2}$ & 0.95$\pm0.05$  & 1.22$_{-0.15}^{+0.16}$ \\
$\chi^2/$d.o.f & 367.3\,/\,387 & 41.5\,/\,59 \\
\noalign{\smallskip}
F$_X$ ($10^{-13}$ erg\,cm$^{-2}$\,s$^{-1}$) & 35.9$\pm1.8 $ &
5.9$_{-1.0}^{+0.7}$ \\
L$_X$ ($10^{35}$ erg\,s$^{-1}$) & 11.2 & 1.8  \\
\noalign{\smallskip}
\hline
\end{tabular}
\tablefoot{We give absorbed fluxes and unabsorbed luminosities in the 0.3--10
keV band. The uncertainties for the spectral parameters are at the 90\,\%
confidence level. The spin periods $P_s$ are given with $1\sigma$
uncertainties (68\,\% confidence level).}
\end{table}

	\subsection{ROSAT observations}
	\label{xray_rosat_observations}
The catalogue of \textit{ROSAT} PSPC sources in the LMC
\citep{1999A&AS..139..277H} includes \object{[HP99] 659}, for which the error
circle contains \LXP. The second \textit{ROSAT} PSPC catalogue has two
detections whose positions are consistent with \LXP\ (\object{2RXP
J050754.4-682457} and \object{2RXP J050756.5-682452}). We searched the
\textit{ROSAT} archive for all observations where \LXP\ was located in the field
of view, at an off-axis angle smaller than 45\arcmin. We found three such
pointings. In two of them (rp500037 and rp300129, those used for the catalogue
entries given above), the source was detected with count rates of
$8.2\pm1.6\times 10^{-3}$ and $3.2\pm1.2\times10^{-3}$ cts\,s$^{-1}$,
respectively. This translates into a 0.3--10 keV flux of \mbox{(4.0 -- 10.2)
$\times 10^{-13}$ erg\,cm$^{-2}$\,s$^{-1}$},  assuming the same spectral
parameters as measured in observation XMM1, which have better statistics. In the
shallower observation rp500060, the source was not detected. We used the
\texttt{ximage} tool \textit{uplimit} to estimate a $3\sigma$ upper limit for
the intensity of $3.1\times10^{-3}$ cts\,s$^{-1}$, corresponding to a 0.3--10
keV flux level $\lesssim 3.9\times10^{-13}$ erg\,cm$^{-2}$\,s$^{-1}$.
\textit{ROSAT} observations and results are summarised in
Table~\ref{table_info}.

\section{The optical counterpart}
\label{optical_counterpart}

\subsection{Identification}
\label{optical_identification}
\begin{figure}[t]
\centering
\includegraphics[bb= 48 142 576
660,clip,width=0.999\hsize]{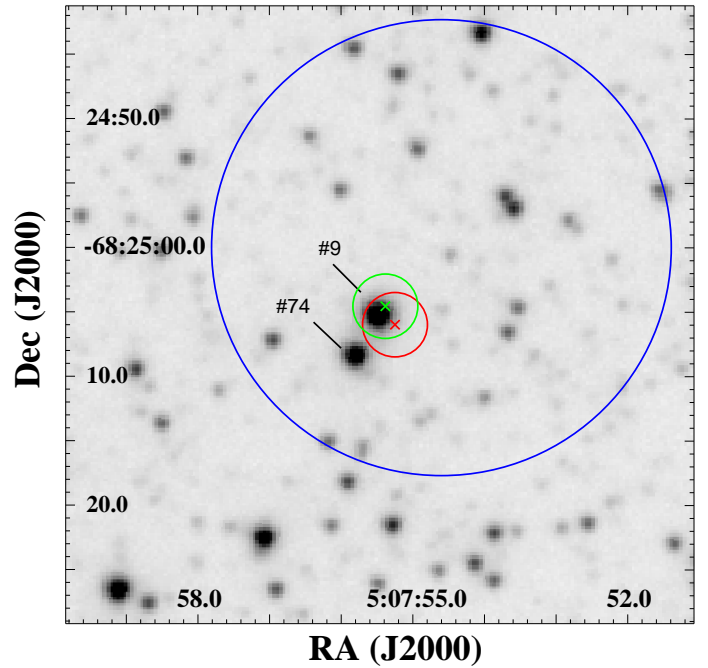}
\caption{OGLE-III $I$-band finding chart for \LXP. The crosses and circles show
the positions of the source and associated $3\sigma$ error circles as measured
in observations XMM1 (green) and XMM2 (red). The large blue circle is the 90\,\%
confidence level error circle of the position of the \textit{ROSAT} source
[HP99] 659. The two possible optical counterparts discussed in the text are
indicated, with star \#9 identified as the true counterpart.}
\label{fig_finding_chart}
\end{figure}

The much better astrometrical accuracy of \xmm\ compared to \textit{ROSAT}
allows us to identify the optical counterpart. We searched for an optical
counterpart to the X-ray source in the catalogue from the Optical Gravitational
Lensing Experiment (OGLE) III survey \citep[using a telescope based at Las
Campanas, Chile, see][]{2008AcA....58...89U}, whose positions have also been
matched to 2MASS. As maximum separation between X-ray and optical positions, we
take the $3\sigma$ uncertainty $r = 3 \sqrt{r_{stat}^2 + r_{syst}^2}$ $\approx
2.5$\arcsec\ for both observations XMM1 and XMM2, as the systematic uncertainty
dominates over the statistical one. We found two candidates near the X-ray
position (see finding chart in Fig.\,\ref{fig_finding_chart}). In the OGLE-III
nomenclature, these stars are lmc.118.4.46009 and lmc.118.4.46074, which we
hereafter refer to as ``OGLE 9'' and ``OGLE 74'', for short. OGLE 9 is the
nearer source, being 0.8\arcsec\ and 1.4\arcsec\ away from the X-ray positions
measured in observations XMM1 and XMM2. i.e. within the $3\sigma$ error circle.
It corresponds to the source \object{2MASS J05075546-6825052}
\citep{2006AJ....131.1163S}. The separations with OGLE 74 are significantly
larger (4.4\arcsec\ and 3.8\arcsec), suggesting OGLE 9 is the true counterpart
to \LXP.

Because we expect the optical counterpart to be an early-type star, most
probably a Be type star, we can secure the association of \LXP\ with OGLE 9,
even in the absence of spectroscopic observations, by studying the
optical and near-infrared (NIR) photometry, colour, and variability of OGLE 9.
In Table\,\ref{table_photometry}, we list the optical photometry of the
two candidates, using the Magellanic Clouds Photometric Survey
\citep[MCPS,][]{2004AJ....128.1606Z} and OGLE-III, and the NIR magnitudes from
the 2MASS and DENIS catalogue. We also include the ultraviolet UVW1 magnitude
measured with \swift\ (see Sect.\,\ref{eclipses_uvot}). The USNO-B1 catalogue
only includes one source located between the two candidates, showing that due to
confusion one cannot use these magnitudes.

\begin{figure*}[t]
	\centering
	\includegraphics[width=0.485\hsize]{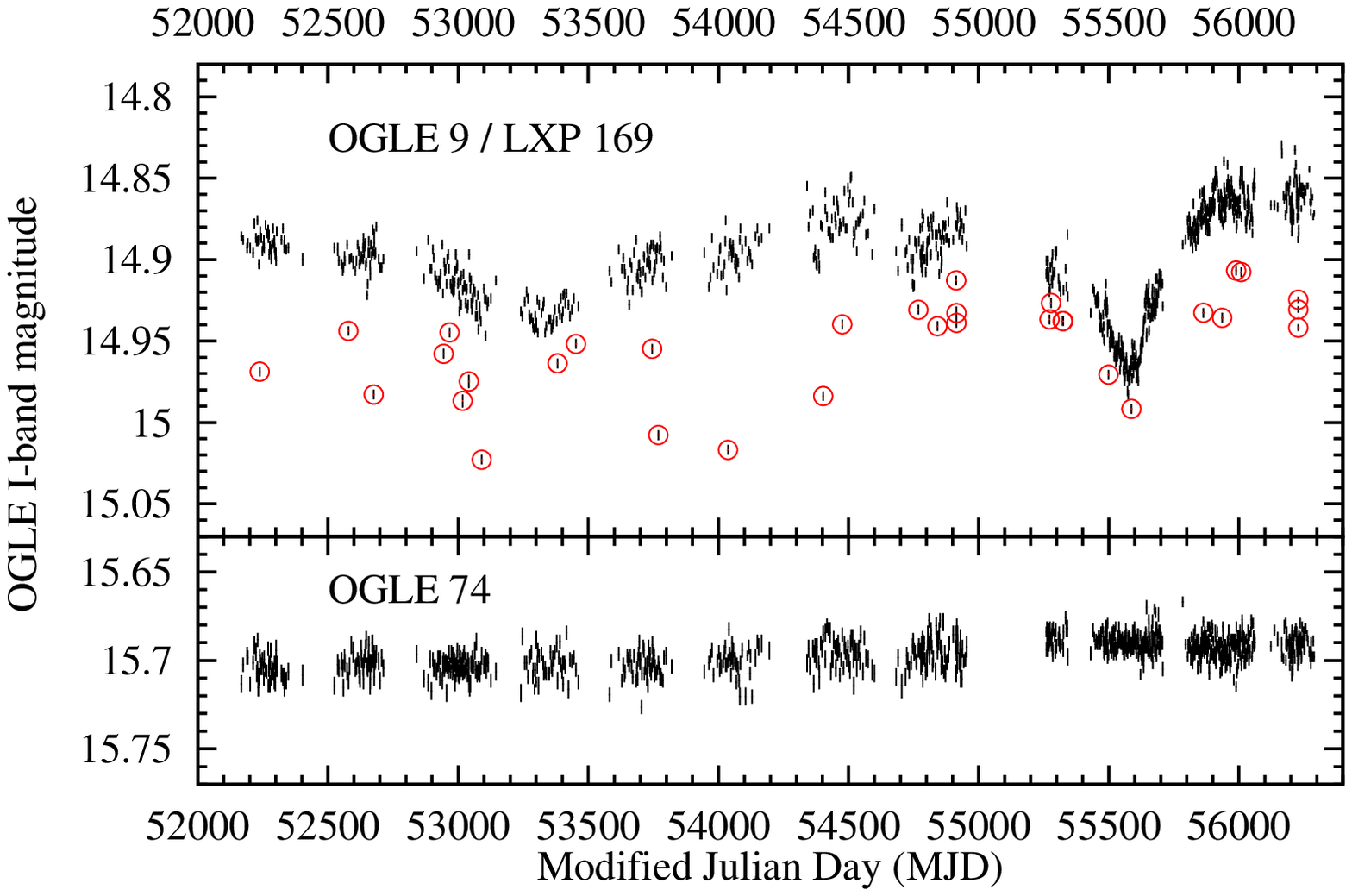}
	\hspace{0.3cm}
	\includegraphics[width=0.485\hsize]{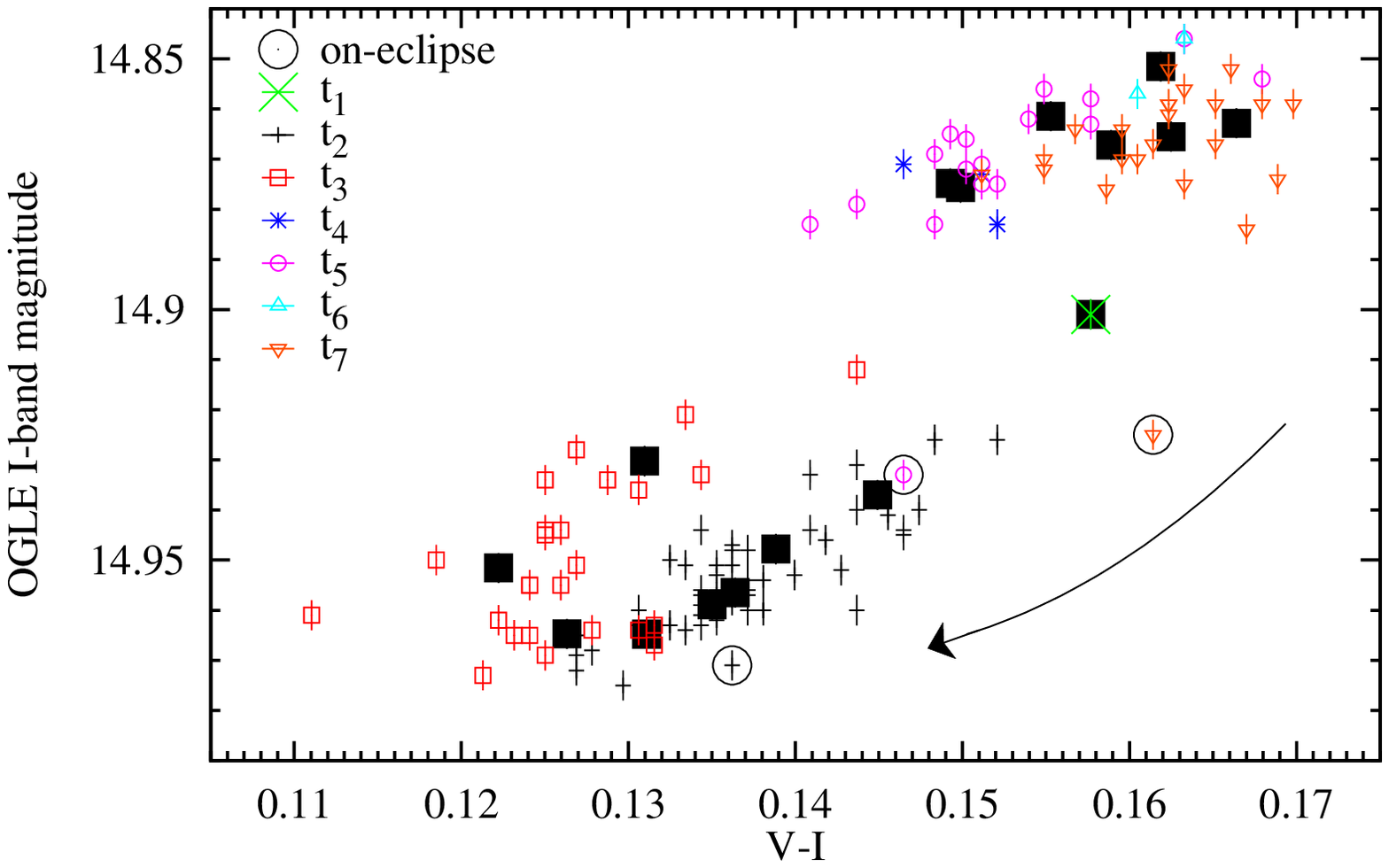}
	\caption{\emph{Left:} OGLE light curves for the two nearest stars from \LXP.
The top panel shows the star we identify as the true optical counterpart (see
Sect.\,\ref{optical_identification}). Data taken before MJD 55000 belong to the
OGLE-III dataset, data taken afterwards are from OGLE-IV. The transit events
(Sect.\,\ref{optical_lightcurve}) are marked by red circles.
\emph{Right:} Colour-magnitude diagram of the optical counterpart of \LXP,
constructed from OGLE-IV $V$ and $I$-band measurements. Data points are
binned in seven 150\,d long blocks (t$_1$ to t$_7$, in chronological order) to
show the path of the star in the diagram. The big filled squares are averaged
values in intervals of about ten consecutive measurements, excluding data
obtained during transit events (circled in black). The curved arrow indicates
how the loop-like track is traversed clock-wise.
}
\label{fig_ogle_lightcurve}
\end{figure*}

The reddening-free $Q$-parameter $Q_{UBV} = (U-B) - 0.72(B-V)$ is -1.049\,mag
and 0.748\,mag for OGLE 9 and 74, respectively. It is derived from MCPS $UBVI$
photometry, which might have been measured non-simultaneously. Although this is
not ideal when we expect variability, it shows that OGLE 9 is of a much earlier
spectral type than OGLE 74. The optical and NIR colours of OGLE 9
($B-V=0.07$\,mag and $J-K=0.42$\,mag) are typical of a Be star
\citep[e.g.][]{2005MNRAS.356..502C}, whereas OGLE 74 appears as a red giant.
Finally, OGLE 9 exhibits a strong optical long-term variability (see
Fig.\,\ref{fig_ogle_lightcurve}, left; details of the light curve are discussed
in Sect.\,\ref{optical_lightcurve}) that is typical of the optical counterpart
to a BeXRB, whereas OGLE 74 does not show any significant variability.

Based on all evidence available (position, photometry, variability), OGLE 9
appears as a typical counterpart to a Be/X-ray binary system. We conclude it is
the true counterpart to \LXP. Spectroscopic observations would enable us to
definitely confirm the emission-line star nature of the counterpart and
determine its spectral type.

\begin{table}[t]
\caption{UV to NIR photometry of the optical counterpart.}
\label{table_photometry}
\centering
\begin{tabular}{c c c c c}
\hline
\hline
\noalign{\smallskip}
Band & \multicolumn{2}{c}{Magnitude} & Catalogue & Ref. \\
 & OGLE 9 & OGLE 74 &  \\
\hline
\noalign{\smallskip}
U & 14.018$\pm$0.080 & 20.317$\pm$0.175 & MCPS & 1 \\
B & 15.016$\pm$0.022 & 18.513$\pm$0.061 & MCPS & 1 \\
V & 14.945$\pm$0.026 & 17.046$\pm$0.035 & MCPS & 1 \\
V & 15.106$\pm$0.017 & 17.097$\pm$0.011 & OGLE-III & 2 \\
I & 14.894$\pm$0.038 & 15.672$\pm$0.186 & MCPS & 1 \\
I & 14.906$\pm$0.023 & 15.702$\pm$0.007 & OGLE-III & 2 \\
I & 14.783$\pm$0.04 & --- & DENIS & 3 \\
J & 14.770$\pm$0.039 & 14.637$\pm$0.069 & 2MASS & 4 \\
K & 14.354$\pm$0.080 & 13.696$\pm$0.054 & 2MASS & 4 \\
\hline
\noalign{\smallskip}
Band & \multicolumn{2}{c}{OGLE 9} & Instrument & Ref. \\
\noalign{\smallskip}
\hline
\noalign{\smallskip}
UVW1 & \multicolumn{2}{c}{14.94$\pm$0.03 } & \textit{Swift/UVOT} & 5 \\
g'   & \multicolumn{2}{c}{14.80$\pm$0.05 } & GROND & 5 \\
r'   & \multicolumn{2}{c}{15.02$\pm$0.04 } & GROND & 5 \\
i'   & \multicolumn{2}{c}{15.21$\pm$0.04 } & GROND & 5 \\
z'   & \multicolumn{2}{c}{15.42$\pm$0.04 } & GROND & 5 \\
J    & \multicolumn{2}{c}{15.72$\pm$0.06 } & GROND & 5 \\
H    & \multicolumn{2}{c}{16.01$\pm$0.07 } & GROND & 5 \\
K    & \multicolumn{2}{c}{16.41$\pm$0.09 } & GROND & 5 \\
\hline
\end{tabular}
\tablefoot{OGLE-III magnitudes are the average of 38 and 378 measurements in the
$V$ and $I$ band, respectively. From \textit{Swift} we give the UVW1 magnitude
(in the AB photometric system), based on an average of six 200 s exposures taken
on MJD 56202.}
\tablebib{(1)~\citet{2004AJ....128.1606Z};
(2) \citet{2008AcA....58...89U};
(3) \citet{2005yCat.2263....0D};
(4) \citet{2006AJ....131.1163S}.
(5) This work.
}
\end{table}

\subsection{GROND photometry}
\label{optical_grond}
We observed OGLE 9 on 2012 October 1 in 33 consecutive four minute observations
with the Gamma-Ray Burst Optical/Near-Infrared Detector (GROND) instrument,
mounted on the ESO/MPG 2.2-m telescope in La Silla, Chile. GROND performs
simultaneous imaging in seven filter bands, from visual
(g$^\prime$r$^\prime$i$^\prime$z$^\prime$ filters) to near-infrafred (NIR, JHK
filters). Technical details on the instrument and its operations are presented
in \citet{2008PASP..120..405G}.

 The data were reduced and analysed with the standard tools and methods
described in \citet{2008ApJ...685..376K}. The g$^\prime$, r$^\prime$,
i$^\prime$, and z$^\prime$ photometric calibration was obtained relative to a
standard star calibrated observation of the field taken on 2012 December 6. The
J, H, and Ks photometry was calibrated against selected 2MASS stars
\citep{2006AJ....131.1163S}. During the 2012 October 1 monitoring, the source
remained constant and the resulting magnitudes are presented in
Table\,\ref{table_photometry}.

To deredden the optical observations, we used the extinction curve given by
\citet{1992ApJ...395..130P}, which can be used for both the Milky Way and
the LMC. The foreground Galactic reddening was set to $E(B-V)_{\mathrm{Gal}} =
0.12$\,mag, or $ A_{V\ \mathrm{Gal}} = 0.37$\,mag, using the map of
\citet{1991A&A...246..231S}. Converting N$_{H\mathrm{\ Gal}} = 5.93 \times
10^{20}\ \mathrm{cm}^{-2}$ into extinction, using the correlation from
\citet{1995A&A...293..889P}, yields a similar value ($A_{V\ \mathrm{Gal}} =
0.33$\,mag). The intrinsic LMC reddening is much harder to estimate: although
the LMC is seen at a moderate inclination angle and is only $\sim$1.7 kpc thick
\citep[e.g.][]{2012AJ....144..106H}, \LXP\ could be in front, within, or behind
the LMC disc. The MCPS reddening
estimator\footnote{\url{http://djuma.as.arizona.edu/~dennis/lmcext.html}} gives
an average extinction of $A_{V}=0.51$\,mag, using all the hot stars within
5\arcmin\ of \LXP, but the scatter is large (standard deviation of 0.42\,mag).
However, we adopted this value, as it is consistent with the best-fit
N$_{H\mathrm{\ LMC}}$ of observation XMM1 which gives $A_{V\ \mathrm{LMC}} =
0.55$\,mag. As a lower limit we assumed no LMC extinction. For the upper limit
we converted the total LMC N$_H$ measured from \ion{H}{I} observations
\citep{2003ApJS..148..473K} at the position of \LXP\ ($2.6  \times 10^{21}\
\mathrm{cm}^{-2}$) into $A_{V\ \mathrm{LMC}}^{\mathrm{max}} = 0.73$\,mag, taking
into account the half-solar metallicity of the LMC.

\begin{figure*}[t]
\centering
\includegraphics[width=\hsize]{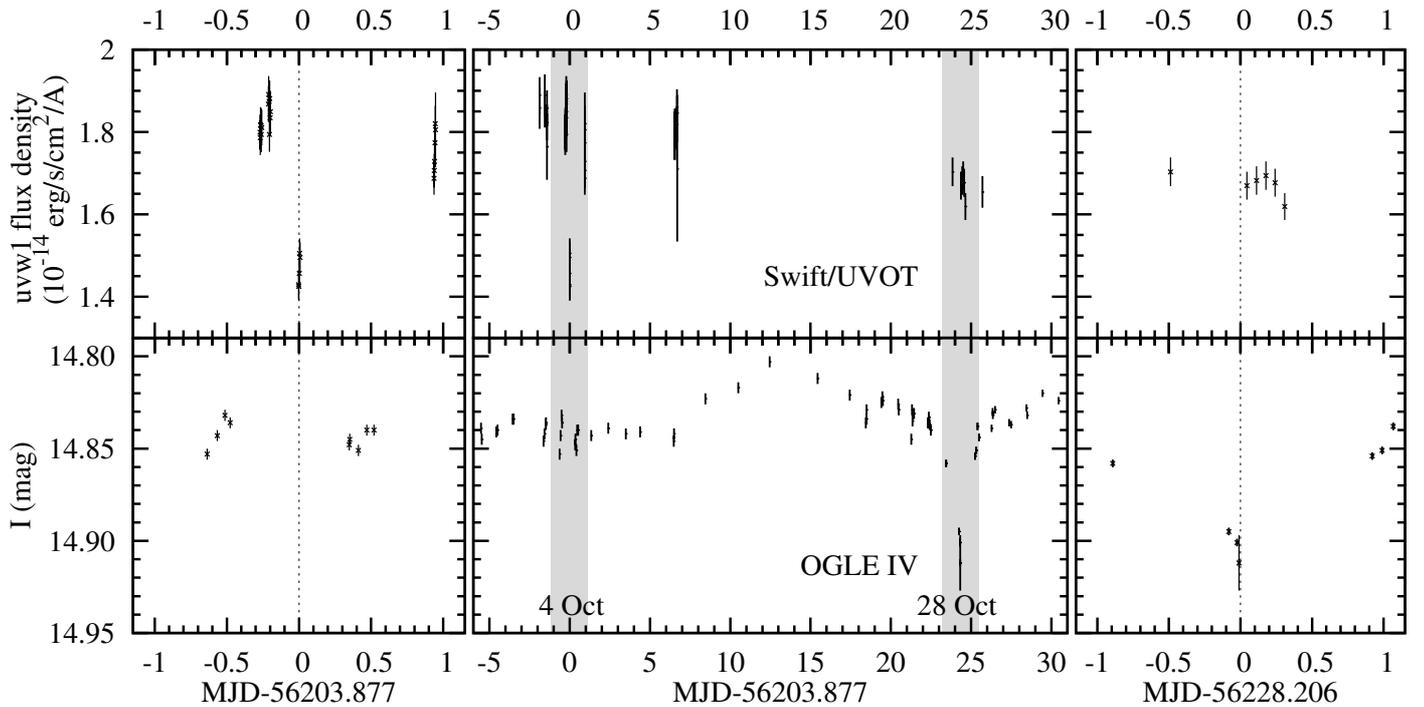}
\caption{Joint UV and $I$ light curve of \LXP\ covering the two eclipses of
October 2012, with data from \swift/UVOT (upper panel) and OGLE-IV (lower
panel). The left and right panels show close-ups of the light curves centred on
the two eclipses, using the transit ephemeris derived in
Sect.\,\ref{eclipses_uvot}. The shaded areas in the middle panel indicate the
zoomed-in regions.}
\label{fig_lightcurve}
\end{figure*}

\subsection{OGLE light curve}
\label{optical_lightcurve}
Combining phase III and IV of the OGLE program, one gets accurate photometry of
the optical counterpart to \LXP, for a period of time longer than 4000\,d on
an almost daily basis (Fig.\,\ref{fig_ogle_lightcurve}, left, top panel). The
light curve is characterised by brightness changes as large as 0.15 mag, typical
of the long term behaviour of Be stars \citep{2005MNRAS.361.1055S}.

We also make use of the OGLE $V$-band observations to study the colour
variability of \LXP. In phase IV of the OGLE program there are 106 simultaneous
(i.e. taken the same night) $V$ and $I$ measurements, which were used to produce
an $I$ vs. $V$-$I$ colour-magnitude diagram (CMD,
Fig.\,\ref{fig_ogle_lightcurve}, right). Correction for colour terms was
performed as described in \citet{2008AcA....58...69U}. Strikingly, the star
follows a loop-like track in the CMD, and the loop is traversed in a clock-wise
sense. This was found to be a rather ubiquitous property of classical Be stars
(at least in the SMC) by \citet{2006A&A...456.1027D}. These authors showed that
this bi-valued colour-magnitude relation can be explained by the presence of an
\emph{outflowing} circumstellar disc and a variable mass-loss rate of the
central star. The loop-like track in the CMD is yet another (strong) evidence
that the companion is indeed a classical Be star. The detection of an infrared
excess with GROND, as presented in Sect.\,\ref{results_discussion_sed}, is
consistent with that picture.

The most intriguing feature of the light curve is the presence of drops in
brightness, appearing at all epochs. These dips are short, as they are usually
seen in only one OGLE datum point at a time (see discussion in
Sect.\,\ref{results_discussion_eclipsing_duration}), and are up to 0.1 \mbox{mag
deep.}

We empirically searched for a periodicity by assuming the interval between the
first and last clear dips (at MJD 52238.25 and 55936.15) to be an integer number
$N$ multiple of a period $P$. We varied $N$ until all dips occurred at their
predicted time. We obtained a satisfactory result for $N=152$, corresponding to
a tentative period of $P=24.328$\,d. We can get a first estimate of the error
on the period by noting that when using $N=151$ or 153, many dips were not
matching the predicted occurrence time. This gives a conservative error for this
tentative period of 0.16\,d.

These periodic events would naturally be explained if the binary system is seen
at high inclination angle. One then expects to observe \emph{eclipses} (or
better here, \emph{transits}, as the flux does not fall to zero) separated in
time by an interval reflecting the orbital period of the system. We present
detailed observations of two recent transits in Sect.\,\ref{eclipses}. Note that
in the remainder of the text we will use both the terms ``eclipse'' and
``transit''.

\section{Dedicated observations of two recent transits}
\label{eclipses}

\subsection{Transits in UV and $I$-band}
\label{eclipses_uvot}
To study \LXP\ and its enigmatic eclipses, we increased the observing frequency
of the system in the OGLE-IV project. When possible, it was followed at least
once per night throughout all the orbital period, and even more often the nights
of the predicted eclipses. Then, we requested observations with the \swift\
observatory. This enables us to measure the visual (OGLE-IV) and UV magnitudes
(using the \swift\ UV/Optical Telescope, UVOT) of the counterpart, while
simultaneously monitoring possible X-ray emission from the neutron star (with
the X-Ray Telescope, XRT, see Sect.\,\ref{eclipses_xrt}).

For the UVOT filter we asked UVW1, firstly because no UV magnitude of the system
existed in the literature, and secondly, because contribution from the
circumstellar disc around the Be star should be minimal in the UV compared to
the $I$-band. We can therefore check if the star itself is eclipsed. In
addition, contaminating light from the close neighbour (in projection) OGLE 74
might in principle fall within the UVOT aperture, but since it appears to be a
red giant, its UVW1 magnitude is at least a factor of 7 or 8 fainter than \LXP.
It should therefore contribute only by a negligible amount.

We were awarded 8.1\,ks of \swift\ observing time, half of it around the
predicted eclipse time MJD 56203.76, as determined using $P=24.328$ d (see
Sect.\,\ref{optical_lightcurve}). The other 4\,ks were distributed before and
after the eclipse, allowing a measurement of the ``out-of-eclipse'' UV
magnitude. We used the HEASoft task UVOTMAGHIST to derive flux densities from
all UVOT exposures. The UVW1 light curve of this observation is shown in the top
panel of Fig.\,\ref{fig_lightcurve}, between MJD 56202 and MJD 56211.

A clear dip is seen at MJD 56203.877 in five UVOT exposures.
These ``on-eclipse'' observations span only 15 minutes in total and we cannot
find any variability amongst them given the flux uncertainties. Nonetheless, we
conclude that \swift\ successfully observed an eclipse of the optical
counterpart. Simultaneous observations of this eclipse with the OGLE telescope
were not possible as it occurred during Chilean daytime. The $I$-band
magnitude measured the nights before and after this eclipse are shown in the
lower left panel of Fig.\,\ref{fig_lightcurve}. A conservative duration limit of
1.13 day for this transit is set by the last \swift\ measurements before and
after the transit. We discuss the transit duration in more details in
Sect.\,\ref{results_discussion_eclipsing_duration}.

With the constraint on the duration of the transit and the $~4000$\,d long
baseline of the OGLE monitoring, we could improve the period estimate to
$P_{\mathrm{orb}}=24.329\pm0.008$\,d. We propose an ephemeris for the
transit of:
\begin{equation}
	\label{eq_ephemeris}
	\mathrm{MJD} \ 56203.877_{-0.197}^{+0.934} + N \times
(24.329\pm0.008)
\end{equation}

We obtained a second set of observations with \swift\ around the time of the
predicted next transit (MJD 56228.206$_{-0.205}^{+0.942}$), using the same
instrumental setting as the first time. We also planned simultaneous optical
follow-up from the ground, using the OGLE telescope and the GROND instrument.
Much to our dismay, thick clouds covered almost all of northern Chile the night
of the eclipse (and the night before). While no image at all could be taken with
GROND, the OGLE telescope managed to take three images through holes in the
clouds (at MJD 56228.18, see Fig.\,\ref{fig_lightcurve}). Compared to previous
and subsequent measurements, the $I$ magnitude was 0.07 mag fainter the night of
the predicted transit, i.e. much larger than the typical photometric accuracy of
the OGLE survey ($\sim$0.003 mag). In these three observations the magnitude was
declining, suggesting that we partially resolved the transit's ingress in $I$.

Although we had quasi-simultaneous observations with \swift/UVOT, we did not
detect a transit as sharp and significant in UV as one orbit earlier. However
the UVW1 flux density was $\sim 10$\% fainter than the ``out-of-eclipse'' level
measured in the first dataset. The observations around MJD 53228.206 span 1.8\,d
in total. A shallower and longer transit could still explain these results (see
Sect.\,\ref{results_discussion_eclipsing_duration}).

	\subsection{Simultaneous X-ray monitoring}
	\label{eclipses_xrt}
The XRT did not detect \LXP\ in the first set of \swift\ observations, around
MJD 56203 (Table\,\ref{table_info} gives observation details). We estimated
$3\sigma$ upper limits of 3 to 8 $\times\ 10^{-13}$ erg\,cm$^{-2}$\,s$^{-1}$ for
the 0.3--10 keV fluxes in the various exposures, using the same method as used
in Sect.\,\ref{xray_rosat_observations}.

In the second data set (around MJD 56228), however, we detected the X-ray source
in all the XRT observations. We measured the count rates and translated them
into 0.3--10 keV fluxes, assuming the spectral parameters found in observation
XMM1 (the uncertainties in the count rate dominate over the uncertainties in the
spectral parameters). We found fluxes ranging from 8 to 11 $\times\ 10^{-13}$
erg\,cm$^{-2}$\,s$^{-1}$. We cannot assess the short-term variability of the
X-ray emission throughout the transit, given the large uncertainties associated
to the flux measurement in such short XRT observations.

\section{Results and discussion}
\label{results_discussion}

	\subsection{Spectral energy distribution}
	\label{results_discussion_sed}

The spectral energy distributions (SEDs) of OBe stars have two components
\citep[e.g.][]{1988A&A...198..200W,1994A&A...290..609D}: a hot stellar
atmosphere with an effective temperature depending on the spectral type (in the
range 18\,000 K -- 38\,000 K), and an infrared (IR) excess produced in a dense
equatorial disc (from which emission lines are also emitted). The presence of
such a disc is supported by the colour-magnitude variability
(Sect.\,\ref{optical_lightcurve}). Because we have the photometry of the
optical companion from NIR to UV, we can study its SED and look for an IR
excess. As Be stars are intrisically variable, their SED
should be constructed from observations taken (quasi-)simultaneously. For this
reason, values from MCPS observations were not included. Instead we took
advantage of the broad-band coverage of the GROND photometry, supplemented by
the UV data from \swift\ (described in Sect.\,\ref{eclipses_uvot}), obtained
less than two days later.

\begin{figure}[t]
	\centering
	\includegraphics[angle=-90,width=0.99\hsize]
{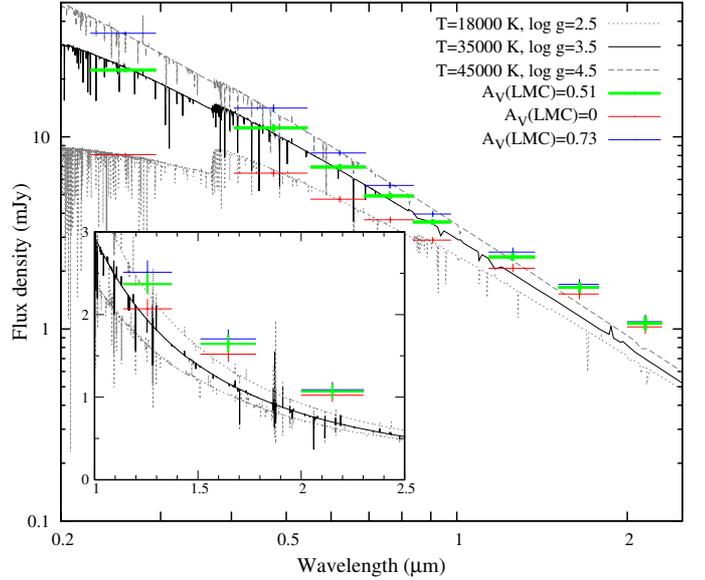}
	\caption{Flux densities of the optical counterpart to \LXP, from UV to NIR
wavelengths. Observations have been dereddened, with an LMC extinction between
0 (red) and 0.73 (blue). Green points show the results for the adopted
extinction of $A_{V\ \mathrm{LMC}} = 0.51$ mag. Stellar atmosphere models of O
and B stars are overlaid (see details in Sect.\,\ref{results_discussion_sed}).
The inset is a close-up view on the JHK bands region, showing the NIR excess.
}
\label{fig_SED}
\end{figure}

The dereddened flux densities are shown in Fig.\,\ref{fig_SED}. We used the
TLUSTY grid of stellar atmosphere models \citep[][for O and B spectral types,
respectively]{2003ApJS..146..417L,2007ApJS..169...83L} to reproduce the
observed photometry. The model normalisations were set by the ratio of the
radius of the star to the distance, for which we took 50 kpc. Depending on the
adopted LMC extinction, temperatures from 18\,000 to 45\,000 K and radii in the
range 8--12 $R_{\sun}$ were required to yield a good agreement of the model
with the data from the UVW1 band to the z' band. In any case, the JHK bands
show a moderate but significant excess, providing strong evidence that the
companion is, indeed, an OBe-type star.

In the case with no LMC absorption, the required temperature would suggest a
spectral type $\sim$ B3 \citep[for luminosity class V,][]{1973AJ.....78..929P},
which is somewhat too late for a typical BeXRB. The spectral types of the
companions in LMC BeXRBs are in the range O9--B2 \citep{2002A&A...385..517N},
though the sample is small compared to the SMC one, where the distribution
extend up to O5 and down to B9 \citep{2005MNRAS.356..502C}. \textit{A
contrario}, in the case with a high LMC extinction, an effective temperature of
45\,000 K would require an O5V type \citep{1996ApJ...460..914V}, which is then
too early (with the previous caveats). For the adopted value of $A_{V\
\mathrm{LMC}} = 0.51$, the effective temperature (35\,000 K) indicates an
O9V-O9.5V type, suitable for a BeXRB. In addition, the predicted $V$-band
magnitude in this case \citep{1996ApJ...460..914V} is the only one to reproduce
the observed value. However, only future spectroscopic observations will confirm
the spectral type.

	\subsection{An eclipsing Be/X-ray binary}
	\label{results_discussion_eclipsing}

The association of the variable X-ray source \LXP, exhibiting pulsations and a
hard spectrum, with the optical counterpart OGLE-III lmc.118.4.46009, a highly
variable early-type star with a near-infrared excess produced in an outflowing
circumstellar disc, leads us to the conclusion that this system is a BeXRB. It
is the 15$^{\mathrm{th}}$ member of the BeXRB LMC population.

What makes this system especially interesting is the \emph{eclipsing} light
curve of the optical counterpart. We now discuss a scenario in which the compact
object transiting in front of its companion causes the features observed in the
light curve of the normal star. We make use of two strong observational results,
namely the depth of the transit seen in UV, and the orbital period
$P_{\mathrm{orb}}$, which is well constrained to $24.329\pm0.008$ d.
In addition, the X-ray pulsations detected with \xmm\ are the spectral
fingerprints of a neutron star, so we can assume the canonical value of
1.4$M_{\sun}$ as the mass of the compact object $M_X$. Other orbital parameters,
such as the inclination $i$, eccentricity $e$, and longitude of periastron
$\omega$ are unknown, although $i$ must be rather close to 90\degr\ for transits
to be observable, and $e$ is usually between 0.3 and 0.9 for BeXRBs
\citep{2011MNRAS.416.1556T}.

		\subsubsection{Size of the eclipsing object}
		\label{results_discussion_eclipsing_size}

Because of its very small size (radius of $\sim$10\,km), it is obvious that a
neutron star will not cause any detectable change in the light curve of a B-type
star when passing in front of it. In this case however, the neutron star orbits
around a star having a dense and slow equatorial outflow. The intense
gravitational field of the neutron star can capture matter up to a radius much
larger than its own surface or magnetospheric radius. The highly enhanced
density in this extended region can block light from the normal star as the
neutron star transits in front of it.

Much like in the case of a transiting exoplanet, it is possible to derive the
\emph{size} of the eclipsing body from the depth of the dip in the light curve
during transit, using:
\begin{equation}
\label{eq_size}
\frac{\Delta F}{F_0} = \frac{R_{X}^2}{R_{C}^2}
\end{equation}
where $F_0$ is the flux of the normal star, $\Delta F$ is the change in flux
during transit, and $R_X$ and $R_C$ are the radii of the eclipsing object and
companion, respectively. Equation~\ref{eq_size} is formally only valid for
completely opaque objects, but can be used to a good approximation if the
eclipses are caused by a region of large optical depth. It is also assumed here
that the two objects are spherical, but more generally the change in flux is a
measure of the fraction of the eclipsed object area covered by the eclipsing
object. We therefore have the opportunity to measure $R_X$ in terms of the
companion star's radius, using $R_X = R_C \sqrt{\Delta F / F_0}$. The NIR
emission of the companion is contaminated\,/\,dominated by contributions from
the equatorial decretion disc, whereas the UV flux should come only from the
stellar surface. Therefore Eq.\,\ref{eq_size} should be applied using the dip
observed with \swift/UVOT on MJD 56203.877, i.e. $\Delta F/F_0 = 0.23$. This
value might be larger because we have not covered a full transit yet. Until
then, and also because of the finite optical depth, we can use $R_X = 0.48 R_C$
as a lower limit. Using a radius suitable for a non-supergiant Be star, for
instance $R_C \sim 10$ \rsun\ means $R_X\gtrsim5$ \rsun.

The observed transit depth also rules out the possibility that \LXP\ is a
triple system, with the eclipsing object being either a planet, a brown
dwarf, or even Sun-like star. The two formers would cause an undetectable change
in flux of $\sim 10^{-4}$ per cent, and the latter a mere 1 per cent.

The incomplete knowledge of the depth of the transit and of the size of the
companion results in large uncertainties in the determination of $R_X$.
Nevertheless, we can perform order-of-magnitude calculations to see if such a
large eclipsing region of dense material around the neutron star can be
formed. We consider the Bondi-Hoyle-Lyttleton accretion process \citep[first
developed by][]{1939PCPS...35..405H,1944MNRAS.104..273B,1952MNRAS.112..195B}: a
neutron
star, going through the Be-star outflow (with density $\rho$) at the relative
velocity $v_{rel}$, can capture matter within the \emph{capture radius} $r_c$
given by \citep{2007rapp.book......}:
\begin{equation}
\label{eq_capture}
r_c = \frac{2 G M_X}{v_{rel} ^2}\ .
\end{equation}
In the simplest case of a circular (or low-eccentricity) orbit, $v_{rel} ^2 =
v_{orb} ^2 + v_w ^2$, with $v_{orb}$ and $v_w$ the orbital and radial outflow
(``wind'') velocities, respectively. Then we have:
\begin{equation}
v_{orb} \approx \sqrt{\frac{G M_C ^2}{a (M_X + M_C)}} ,
\end{equation}
where $M_C$ is the mass of the companion and $a$ the semi-major axis of the
binary. Using Kepler's third law and defining \mbox{$q = M_C / M_X$} we have
\begin{equation}
\label{eq_a}
a = 39.55 
\left[\frac{M_X}{1.4 M_{\sun}}\right] ^{1/3}
\left[\frac{P_{orb}}{24.329 \mathrm{d}}\right] ^{2/3}
\left( 1 + q \right) ^{1/3} 
\ R_{\sun}
\end{equation}
and therefore
\begin{equation}
v_{orb} \approx (2 \pi G) ^{1/3} \ P_{orb} ^{-1/3} \ M_X ^{1/3} \
\frac{q}{(1+q)^{2/3}}
\end{equation}
\begin{equation}
\label{eq_vorb}
v_{orb} \approx 82.2 \ \left[\frac{P_{orb}}{24.329 \mathrm{d}}\right]^{-1/3} \
\left[\frac{M_X}{1.4 M_{\sun}}\right] ^{1/3} \frac{q}{(1+q)^{2/3}} \
\mathrm{km\,s}^{-1} .
\end{equation}
Combining Eqs.\,\ref{eq_capture} \& \ref{eq_vorb} with the appropriate scaling
factors, one obtains:
\begin{equation}
\label{eq_rc}
r_c = 
\frac{ 53.2
\left[\frac{M_X}{1.4 M_{\sun}}\right]
\left[\frac{v_w}{100 \ \mathrm{km\,s}^{-1}}\right] ^{-2}
}
{
1+1.48
\left[\frac{v_w}{100 \ \mathrm{km\,s}^{-1}}\right] ^{-2}
\left[\frac{P_{orb}}{24.329 \mathrm{d}}\right] ^{-2/3}
\left[\frac{M_X}{1.4 M_{\sun}}\right] ^{2/3} 
\frac{q^2}{(1+q)^{4/3}}
}
\ R_{\sun}
\end{equation}
For the wind velocity we use $v_w = v_0 (r/R_{\ast}) ^{n-2}$
\citep{1988A&A...198..200W}, where $v_0$ is the initial velocity at the stellar
surface ($r=R_{\ast}$). It is clear that the uncertainty in estimating $r_c$ is
dominated by our poor knowledge of the wind velocity structure and of the mass
of the companion. We used $v_0 = 0.67$ km\,s$^{-1}$ and $n = 3.5$
\citep{1995A&A...300..259V} and assumed a companion mass of 10\,$M_{\sun}$,
yielding $a \sim 80 \, R_{\sun}$ and $v_w \sim$ 15 km\,s$^{-1}$. Under these
crude assumptions we get $r_c \sim 11.5 \, R_{\sun}$. This is smaller than the
Roche lobe radius of the neutron star $R_{L,X} \sim 18 \, R_{\sun}$ for $M_C =
10 \, M_{\sun}$ \citep[from Eq.\,\ref{eq_a} and the approximation
of][]{1983ApJ...268..368E}, which is the maximal radius up to which the neutron
star can capture matter. Note that $r_c$ is in turn an upper limit for $R_X$,
as the size of the eclipsing object depends on the way the captured matter
evolves around and is accreted onto the neutron star.

The exact effect of a significant eccentricity is difficult to assess: the
neutron star would reach \emph{faster} orbital velocity $v_{orb}$ closer to the
Be star, where $v_w$ is \emph{slower}. Although $r_c$ would become strongly
orbital-phase-dependent, it should remain roughly in the same range as in the
case of a circular orbit.

We also consider, instead of an outflowing wind disc scenario \textit{\`a la}
\citet{1988A&A...198..200W}, a viscously driven decretion disc, which can
explain most of the observations of Be star discs \citep[and references
therein]{1991MNRAS.250..432L,1999A&A...348..512P}. Equation \ref{eq_rc} does
not apply in such a case, because the orbital velocity of the neutron star
matches that of the (quasi-)Keplerian disc. This leads to slow relative
velocities (the radial drift velocities in viscous discs are less than a few
km\,s$^{-1}$) and therefore high values of $r_c$ in the Bondi-Hoyle-Lyttleton
formalism. In that case the Roche lobe radius of the neutron star is most likely
the upper limit for the size of the eclipsing object.

The analysis above shows that, at least to a first order, it is possible that
the transits observed in the light curve of the companion star are caused by the
matter captured by the orbiting neutron star.

		\subsubsection{Transit duration}
		\label{results_discussion_eclipsing_duration}

The transit duration is another proxy for the size of the eclipsing object.
It depends in a complex manner on the orbital parameters (particularly $i$, $e$,
and $\omega$), most of which are yet unknown. In the simple case of a circular
orbit, the duration of transit $T_{circ}$ is
\begin{equation}
\label{eq_transit}
T_{circ}
=
\frac{P_{orb}}{\pi}
\arcsin
	\left(
	\frac{
		\sqrt{
			\left( R_C + R_X \right) ^2 - a^2 \cos^2{i}
			}
		}
		{a}
	\right)
\end{equation}
If we further assume that $i \gtrsim 75 \degr$ we can neglect $\cos^2{i}$.
Then, assuming $R_C = 10 \ R_{\sun}$, $R_X \sim 5 \ R_{\sun}$
(Sect.\,\ref{results_discussion_eclipsing_size}), and $M_C = 10 \ M_{\sun}$,
Eqs.\,\ref{eq_a} \&\ \ref{eq_transit} give $T_{circ} = 1.47$ d, which is
slightly longer than the upper limit of 1.13 d found in
Sect.\,\ref{eclipses_uvot}.

This discrepancy can easily be reconciled by using various reasonable values for
$R_X$, $R_C$, $q$, or adding eccentricity \citep[expected in the case of
BeXRB,][]{2011MNRAS.416.1556T}. For instance, only changing $e$ in the range
0.25--0.5 (leaving other parameters unchanged and using $\omega=0$), yields
transit durations ranging from 1.13 d to 0.85 d). However, there is a high level
of degeneracy between all the parameters and one would first need to measure
some of them (e.g. $R_C$ by spectroscopy, $e$ and $\omega$ by radial
velocity/light curve modelling) before being able to obtain $R_X$ from the
transit duration.

Due to the detection of several \emph{consecutive} transits in the OGLE light
curve, which has a sampling period of typically one day or more, we can estimate
the duration of the transit in the $I$-band. Were these significantly shorter
than a day, one could not easily detect two transits separated by 24.329 d
from the same location (here, Chile), simply because the next one would occur
during daytime. OGLE observations on MJD 54914.03, 54914.99, and 54915.09,
showing the star eclipsed (0.06 mag depth) in two consecutive nights, confirm
that $\sim 1$ day-long $I$-band transits do exist. The absence of transits seen
in three consecutive nights suggest that two days is a reasonable upper limit
for the transit duration in the near-infrared.

We stress, however, that the transit durations are likely to be variable.
Indeed, due to the Be-star activity, the circumstellar disc can grow or shrink,
or even present asymmetries \citep[e.g.][]{1991PASJ...43...75O}, changing the
starting and/or ending point of the transit and resulting in different
durations, in particular in the near-infrared. Asymmetric transits do exist, as
exemplified by OGLE-III $I$-band observations around the transits predicted on
MJD 55279.375 and MJD 55328.033 (using Eq.\,\ref{eq_ephemeris}): 0.03 mag deep
transits were detected, starting six and five days \emph{earlier} than
expected, but not extending after the predicted times. Changes in the velocity
and density structure of the disc would cause the size of the eclipsing object
$R_X$ to vary (as the accretion radius and rate would be affected), resulting in
a shorter or longer transit duration in the UV as well.

		\subsubsection{Comparison to other eclipsing X-ray binaries}
		\label{results_discussion_eclipsing_comparison}
Although a few other eclipsing X-ray pulsar binaries are known \citep[eleven at
the time of the writing,][]{2012MNRAS.422..199M}, they are all supergiant X-ray
binaries (SGXBs) or low-mass X-ray binaries, in which the compact object is
eclipsed by the mass donor. \LXP\ would be the first confirmed eclipsing BeXRB.

Furthermore, we see transits/eclipses \emph{of the optical counterpart} and not
from the X-ray source, as is the case in known eclipsing SGXBs. Such X-ray
eclipses are likely to be also present in \LXP. However, due to
\emph{i)} the transient nature of the X-ray emission in this system;
\emph{ii)} the wider, most likely eccentric orbit; and
\emph{iii)} the absence of an orbital solution,
observing an eclipse in X-ray might prove to be an arduous task.

A similar system is possibly \object{XMMU J010743.1-715933} in the SMC
\citep{2012MNRAS.424..282C}. The light curve of the optical counterpart (their
Fig.\,15) folded to the 100.3 d orbital period of the system reveals
eclipse-like features of $\sim$ 0.1 mag depth ($\Delta F/F_0 \sim 0.09$). Also
intriguing is the fact that these transits are followed by optical outbursts.
These features are not explained in \citet{2012MNRAS.424..282C}, although they
remark the similarity with other transiting eclipsing binaries in the LMC
\citep{2011AcA....61..103G}, for which variations of eclipse features are
explained by regression of the nodes of the orbital plane due to the presence of
a third body in the system.

	\subsection{X-ray variability}
	\label{results_discussion_xray_variability}
The X-ray luminosity of \LXP\ in the available observations (up to $10^{36}$
erg\,s$^{-1}$) suggests we see Type I outbursts, when the neutron star is at or
near periastron. Because the major axis of the orbit does not have to be aligned
with our line-of-sight, there is no reason for the periastron passage and
transit to be simultaneous. The highest X-ray flux observed so far ($36 \times
10^{-13}$ erg\,cm$^{-2}$\,s$^{-1}$) is reached in the observation XMM1, taken at
an orbital phase $0.60 \leq \Phi \leq 0.61$. A flux about six times lower is
measured in observation XMM2, at a phase 0.98--0.99.

\textit{ROSAT} has a much lower effective area and therefore lower accuracy in
the determination of a flux in the $10^{-13}$ to $10^{-12}$
erg\,cm$^{-2}$\,s$^{-1}$ range. The three available \textit{ROSAT} observations
have been done at similar epochs and span orbital phases of 0.43 to 0.66, during
which the flux varied between $10.5 \times 10^{-13}$ erg\,cm$^{-2}$\,s$^{-1}$
and a non-detection, with an upper limit of $\lesssim 4\times10^{-13}$
erg\,cm$^{-2}$\,s$^{-1}$. 

Also \swift/XRT only provides rough estimates of the X-ray flux because of
the short exposure times. While the source was not detected around the transit
on MJD 56203.877 (the flux was always lower than $\sim 8 \times10^{-13}$
erg\,cm$^{-2}$\,s$^{-1}$), observations during the next transit detected the
source at a roughly constant level of 8 to 11 $\times\ 10^{-13}$
erg\,cm$^{-2}$\,s$^{-1}$, i.e. consistent with the flux measured in observation
XMM2, which was performed at a similar orbital phase.

The variability factor is at least 10 but is poorly constrained due to the
absence of high-sensitivity and/or deep observations during ``quiescence''.
Monitoring the X-ray emission during several orbits, across all phases, will
allow us to learn more about the X-ray cycle, such as the phase of the outbursts
onset and their durations. We can also look for a Type II (i.e. longer and
brighter) outburst, in which case an X-ray eclipse would be easier to find.

	\subsection{Future observations}
	\label{results_discussion_future}

The detection of transits in the light curve of the optical counterpart gives
us the unique opportunity to estimate the size and structure of both the matter
captured by the neutron star and the equatorial disc around the Be-star
companion. To do this one needs to cover the transit from NIR to UV wavelengths,
measuring the time of ingress/egress and the transit duration across a broad
spectral domain. If the eclipsing object is indeed completely opaque, the I-band
transit durations are expected to be longer than UV transits, because the
neutron star will first cover one side of the circumstellar disc before passing
in front of the star itself. It will then cover the other side of the disc. If,
on the other hand, transits from UV to NIR have similar durations, it would mean
the eclipsing object has a high but finite optical depth. The latter could be
estimated by measuring colour changes across a transit.

 We already attempted such simultaneous observations using \swift, GROND, and
OGLE during the transit on MJD 56228.18, but bad weather conditions prevented us
from covering the transit (Sect.\,\ref{eclipses_uvot}). The relative short
orbital period of 24.329\,d means we can repeat these attempts more regularly
than for longer period systems, such as \object{XMMU J010743.1-715933} (see
Sect.\,\ref{results_discussion_eclipsing_comparison}), having a period of
100.3\,d.

Optical spectroscopy will provide us with a lot of valuable information.
Firstly, one can confirm the emission-line nature of the companion (as expected
from the observed NIR excess and variability). Secondly, the resulting spectral
classification would provide additional constraints on the size of the Be-star.
Thirdly,  as we know that the orbital plane is highly inclined, we could use the
H$\alpha$-line profile to study a possible Be disc/orbital plane misalignment,
which have been recently suggested by \citet{2012arXiv1211.5225O} to explain the
two types of X-ray outbursts seen in BeXRBs.
If the Be disc were also highly inclined, we expect a double-peaked H$\alpha$
line, and the separation between the blue- and redshifted part of the line
becomes narrower as the disc is seen from a more polar direction (i.e. as the
tilt angle between the orbital plane and the Be disc plane increases).
Another proxy for the Be disc inclination could come from the study of
brightness and colour variations, as observed in
Fig.\,\ref{fig_ogle_lightcurve}. \citet{2012ApJ...756..156H} showed that
inclination angle $i$ is a parameter relevant to the observed photometric
variations and in-depth modelling  of the system, though beyond the scope of
this paper, could result in an estimate for $i$.
It is also worth mentionning that phase-resolved spectroscopy might detect the
equivalent of a Rossiter-McLaughlin effect in the H$\alpha$-line profile. The
main idea is that as the eclipsing neutron star transits in front of the
approaching part of the equatorial disc, it covers first the blueshifted peak of
the line; then, it passes in front of the receding part of the line and the
redshifted peak is masked. This, of course, assumes we see a double-peaked
H$\alpha$ line out of transit.

Ultimately, the \emph{mass} of the neutron star could be determined, as the
eclipsing nature of the system constrains the inclination angle. This requires
to obtain the radial velocity curve of the companion and to measure the pulse
arrival time delays at various phases. The observational effort needed is
justified by the scarcity of dynamically-determined neutron star masses,
even though there are the main constraints on the variety of equations
of state available \citep[e.g.][]{2010arXiv1011.4291K}. A mass has been derived
from only seven out of eleven eclipsing X-ray pulsars
\citep{2011ApJ...730...25R,2012MNRAS.422..199M}.

	\section{Summary}
	\label{summary}
We discovered a new X-ray pulsar in the LMC, using \xmm\ observations. The
neutron star has a spin period of \mbox{168.8 s} and we gave the identifier
\LXP\ for the source. Its spectrum could be characterised by a power law, which
was harder when the flux was higher. The X-ray brightness is variable by a
factor of at least 10, as shown by comparison of the \xmm\ results with archival
\textit{ROSAT} data and subsequent \swift\ follow-up observations.

Using the \xmm\ position, we identified the optical counterpart of \LXP. A
strong variability is present in the long-term OGLE light curve of the
companion, and its path in the colour-magnitude diagram is typical of a star
surrounded by an outflowing circumstellar disc. Observations from UV to NIR
wavelengths revealed an NIR excess, which is most likely produced in such a
disc. These photometric properties are very strong evidence that the companion
is an OBe star. Therefore, \LXP\ is classified as a new BeXRB.

The OGLE light curve suggests that the system is eclipsing. We observed the
system with \swift\ at a predicted eclipse time. The detection of a transit in
the UV proved that the companion itself is eclipsed. We determined an ephemeris
for the transit times of MJD $56203.877_{-0.197}^{+0.934} + N \times
(24.329\pm0.008)$. We proposed that the orbiting neutron star, and in particular
the matter captured from the companion, are responsible for the eclipses. We
used the depth and duration of the transit to derive first constraints on the
size of the eclipsing object.

We concluded that not only is \LXP\ the first confirmed eclipsing BeXRB, but
also that for the first time the companion star in an X-ray binary itself is
seen eclipsed by the compact object. We showed how much can be learned from
future observations of this important system.

\begin{acknowledgements}
We thank the anonymous referee for fruitful suggestions.
The \xmm\ project is supported by the Bundesministerium f\"ur Wirtschaft und
Technologie\,/\,Deutsches Zentrum f\"ur Luft- und Raumfahrt (BMWi/DLR, FKZ 50 OX
0001) and the Max-Planck Society. The OGLE project has received funding from the
European Research Council under the European Community's Seventh Framework
Programme (FP7/2007-2013)\,/\,ERC grant agreement no.\,246678 to A.\,U. P.\,M.
and R.\,S. acknowledge support from the BMWi/DLR grants FKZ 50 OR 1201 and FKZ
50 OR 0907, respectively. Part of the funding for GROND (both hardware and
staff) was generously granted from the Leibniz-Prize to Prof. G.
Hasinger (DFG grant HA 1850/28-1). This research has made use of Aladin, SIMBAD
and VizieR, operated at the CDS, Strasbourg, France. We thank the \swift\ team
for accepting and carefully scheduling the target of opportunity observations,
and we acknowledge the use of public data from the \swift\ data archive.
\end{acknowledgements}


\newpage

\end{document}